\documentclass[%
 reprint,
 amsmath,amssymb,
 aps,
]{revtex4}

\usepackage{graphicx}
\usepackage{dcolumn}
\usepackage{bm}
\usepackage{siunitx}
\usepackage{adjustbox}
\usepackage{graphicx}
\usepackage{float}
\restylefloat{table}
\usepackage{placeins}

\begin{document}

\preprint{APS/123-QED}

\title{Lamb shift in hydrogen-like muonic atoms to test fractal extra dimensions}

\author{Daniela Grandon}
\email{daniela.grandon@alumnos.uv.cl}
 
\author{Alfredo Vega}%
\email{alfredo.vega@uv.cl}
\affiliation{%
 Instituto de F\'isica y Astronom\'ia, \\
 Universidad de Valpara\'iso,\\
 Av. Gran Breta\~na 1111, Valpara\'iso, Chile
}%

\begin{abstract}

From the comparison between the experimental and theoretical results for the energy difference between $2s_{1/2}$ and $2p_{1/2}$ states in hydrogen and hydrogen-like ions with different values of Z, we estimate bounds for the presence of an extra fractional dimension. These bounds are higher when we work with muonic hydrogen and muonic deuterium,  which open the possibility that future measurements of the Lamb shift in muonic atoms with one electron and higher values of Z may show evidence of the extra dimension. 
\end{abstract}


\date{\today}



\maketitle


\section{\label{sec:level1} Introduction}


Even before Pohl et al. published their precise measurements using spectroscopy on muonic atoms that gave origin to the proton radius puzzle \cite{Pohl:2010zza}, muonic atoms had already attracted the interest of the scientific community. These atoms make it possible to verify the standard model and to explore physics beyond it \cite{Hughes, Indelicato:2004sz}. An example of the latter can be seen in \cite{Luo:2006ck, Luo:2006ad}, where the authors suggested the possibility of using these kinds of atoms to detect extra dimensions. After the emergence of the proton radius puzzle, based on a disagreement between the values of the proton radius obtained by elastic collisions e-p using hydrogen and the experiments on muonic hydrogen using spectroscopy, the interest in muonic atoms increased, and since then different alternatives have emerged to explain the discrepancy in proton radius.



The theoretical efforts aimed at solving this problem have included the use of the standard model (SM) as well as physics beyond this model. In this way, research on muonic atoms may open a window to the study of new physics \cite{Pohl:2013yb, Krauth:2017ijq} with experiments where the most important thing is not the increase in energy, but rather the precision in each measurement \cite{Liu:2017bzj, Signer}.



Among the different attempts to solve the proton radio puzzle using physics beyond the SM, several authors have studied models with extra dimensions \cite{Wang:2013fma, Onofrio:2013fea, ONOFRIO:2014eaa, Wan-Ping:2014ela, Dahia:2015xxa, Dahia:2015bza}. As an example, the work proposed by Arkani-Hamed, Dimoupoulos and Dvali \cite{ArkaniHamed:1998rs, Antoniadis:1998ig, ArkaniHamed:1998vp} consider a modified gravitational potential that gives origin to a contribution capable of modify atomic spectrum along the extra dimension. This additional contribution has been suggested as a possible explanation to the discrepancy between the theoretical value of proton radius suggested by CODATA and the measured value obtained in muonic atoms \cite{Wang:2013fma, Onofrio:2013fea, ONOFRIO:2014eaa, Wan-Ping:2014ela, Dahia:2015xxa, Dahia:2015bza}.


The above is not the only way to include extra dimensions in order to introduce corrections to the spectrum of hydrogen atoms. In the mid-1980s, Schafer and Muller published two works \cite{Muller:1985bt, Schafer:1986oda} where they explored the effects of considering extra dimensions upon derivatives of functions, then modifying the Laplacian and the different terms in the Schrodinger equation. When considering a $d+\varepsilon$ dimensional space with $\varepsilon$ small, the Hamiltonian is modified, and using the Hellmann-Feynmann theorem, an additional term to the potential is obtained, which is linear with the fractional extra dimension. On the other hand, the modified derivatives change Gauss’s law and so change the Coulomb type potential. However, this does not produce contributions to the energy shift between levels $2s$ and $2p$ in the hydrogen spectra \cite{Muller:1985bt, Schafer:1986oda}.



From the corrections that this extra dimension has on the Lamb shift, authors in \cite{Muller:1985bt, Schafer:1986oda} obtained upper bounds for the value of the extra fractional dimension according to the theoretical and experimental values available in that era for the $2s$ and $2p$ transitions. These works were considered by other authors, such as \cite{Buettner:1999dv, Buettner:2000rq, Buettner:2003qz}, who explored the early universe using Lyman $\alpha$ and Lyman $\beta$ lines from quasars at high redshift. In this work, we use the formalism of \cite{Muller:1985bt, Schafer:1986oda} to show that bounds obtained for hydrogen-like ions increases as Z does, and these bounds get even higher when we work with muonic atoms, like muonic deuterium. This enables us to suppose that future measurements of the Lamb shift in hydrogen-like muonic atoms with high values for Z may provide more important clues about extra dimensions. 


In \cite{Muller:1985bt, Schafer:1986oda} the authors studied hydrogen obtaining a bound of $10^{-11}$ for the extra fractional dimension. As we will show in this work, taking into account measurements of the Lamb shift in hydrogen-like ions with high values for Z \cite{Yerokhin}, it is possible to estimate new bounds for the extra dimension. 


On the other hand, we consider the equations of \cite{Muller:1985bt, Schafer:1986oda} applied to muonic hydrogen, and the measurements of \cite{Pohl:2010zza}. The result is an upper bound of $\sim 10^{-7}$. This value added to the increasing behaviour of the bounds when Z increases motivates future measurements of the Lamb shift in muonic atoms with high Z as projected by CREMA in order to get the highest values for the extra dimension.

This work is organized as follows. In section II we present the general ideas and the results taken from \cite{Muller:1985bt, Schafer:1986oda} that will be used to estimate bounds for the fractional extra dimension. Our results are also included in that section, using experimental and theoretical values for the the Lamb shift in hydrogen-like ions and muonic atoms. Section III presents the conclusions and motivates new experiments in this area. 





\section{Calculation of bounds for fractional extra dimension}

In \cite{Muller:1985bt, Schafer:1986oda} the authors considered a generalization of the Schrodinger equation, with a $d$-dimensional Laplacian. If a dimension such as $d = 3 + \varepsilon$ is used, where $\varepsilon$ is a small extra fractionary dimension, an equation similar to the three - dimensional one is obtained, where the effect of this extra dimension is absorbed into a term added to the potential, which depends explicitly on this extra dimension's parameter. This new term on the potential can be treated by perturbation theory due to the small size of $\varepsilon$. Gauss's law and Coulomb's potential are also modified but they do not generate a shift in the energy for the transitions we are studying. 



From the mentioned generalization, we obtain \cite{Muller:1985bt, Schafer:1986oda}


\begin{equation}
\label{Ec1}
\Delta E_{LS}=E_{3+\varepsilon}^{2p_{1/2}}-E_{3+\varepsilon}^{2s_{1/2}}\approx- \frac{(Z\alpha)^2 \mu}{12}\varepsilon,
\end{equation}
where the theoretical values for levels $2s$ and $2p$ contain contributions taken from the standard model calculations plus the contribution resulting from the extra dimension, so the above expression can be written as  

\begin{equation}
\label{Ec}
\Delta E_{LS}=\Delta E_{LS}^{teo}-\Delta E_{LS}^{exp}\approx- \frac{(Z\alpha)^2 \mu}{12}\varepsilon
\end{equation}
where $\Delta E_{LS}^{teo}$ is the theoretical value using physics from standard model, $\alpha$ is the fine structure term and $\mu$ the reduced mass of the atomic system we are studying.


In the case of the hydrogen atom, expression (\ref{Ec1}) is reduced to 
\begin{equation}
\label{DeltaELShidrogeno}
\Delta E_{LS} \approx - 2.27~\varepsilon~eV.
\end{equation}

Using measurements for the $2p_{1/2}-2s_{1/2}$ transition in hydrogen-like ions with different Z \cite{Yerokhin}, equation (\ref{Ec}) enables us to get different upper bounds for $\varepsilon$. Table I shows the results for the fractional extra dimension using the measurements from the Table 5 of \cite{Yerokhin} for different hydrogen-like ions with only one electron (where the last column includes the references to the experimental values of the Lamb shift in each different case of Z).
This table clearly shows an increasing behavior for $\varepsilon$ as the atomic number grows, i.e., the effect of this extra dimension gains in importance as the nucleus becomes heavier.


For the case of hydrogen ions, we consider the electron mass as the reduced mass of the system; however, for atoms where the electron is replaced by a muon, the reduced mass changes. Therefore we are going to rewrite equation (\ref{DeltaELShidrogeno}) to be applied easily in muonic hydrogen with an arbitrary number of nucleons $A$. Thus, the reduced mass can be written as


\begin{equation}
\mu=\frac{m_\mu A m_p}{m_\mu + A m_p}= \frac{m_\mu}{\left(\frac{m_\mu}{A m_p}\right)+1}= f_{Z,A} m_\mu
\end{equation}
where $m_\mu$ is the mass of the muon and $f_{Z,A}=(1+\frac{m_\mu}{A m_p})^{-1}$ is a linear function of A. In particular using $m_\mu=105.65$ MeV and $m_p=938.2$ MeV for muonic hydrogen we obtain $f_{Z,A}=0.9$ and


\begin{equation}
\label{DeltaMuonico}
\Delta E_{LS}\approx-469.08~Z^2~f_{Z,A}~\varepsilon.
\end{equation}

Using the theoretical expressions for the Lamb shift, with the proton radius $0.8751(61)$ fm given by CODATA-2014, and the experimental measurements from \cite{Pohl:2010zza} for muonic hydrogen we get $\Delta E_{LS} \approx 0.296(56)$ meV. Thus


\begin{equation}
\varepsilon=\num{7.013e-7}\pm \num{1.3e-9},
\end{equation}
Muonic deuterium \cite{Pohl1:2016xoo} gives $\Delta E_{LS} \approx 0.409(66)$ meV, and with the deuteron radius $2.1413(25)$ fm given by CODATA-2014, leads to a bound for the extra dimension of 

\begin{equation}
\varepsilon=\num{9.203e-7} \pm \num{1.5e-9}.
\end{equation}

We can see a slight increase in the bound for the extra dimension with the number of nucleons, which is consistent with the results shown in Table I for hydrogen ions with electrons. Therefore, we interpret this as atoms with low $ Z $ not being good atomic systems to detect extra dimensions.


\begin{table*}[t]
\begin{center}
\begin{tabular}{| c | c | c | c | c | c | c |}
\hline
$Z$ & Element & Theory (eV) & \multicolumn{3}{ c |}{Experiment} & $\varepsilon$ \\ 
\cline{4-6}
 &  &  & Result (eV) & Year & Ref & \\
\hline
\hline

~~1~~ & ~~H~~ & ~~0.00000437495(5)~ & ~~0.000004374893(1)~~ & ~~1981~~ & ~~\cite{Lundeen:1981zz}~~ & ~~$\num{3.63e-11} \pm \num{2.2e-12}$~~ \\ 
 \hline
~~2~~ & ~~$^{4}$H~~ & ~~0.0000580708(1)~~ & ~~0.0000580694(7)~~ & ~~2000~~ & ~~\cite{vanwij:2000van}~~ & ~~$\num{1.55e-10} \pm \num{7.8e-12}$~~ \\ 
  \hline
~~3~~ & ~~$^{6}$Li~~ & ~~0.00025945(1)~~ & ~~0.00025958(9)~~ & ~~1975~~ & ~~\cite{leventhal:1975lev}~~ & ~~$\num{6.39e-9}\pm \num{4.4e-10}$~~ \\ 
 \hline
~~16~~ & ~~$^{32}$S~~ & ~~0.104883(3)~~ & ~~0.10449(26)~~ & ~~1993~~ & ~~\cite{brentano:1993bren}~~ & ~~$\num{6.77e-7}\pm \num{4.5e-9}$~~ \\ 
 \hline
~~17~~ & ~~$^{35}$Cl~~ & ~~0.12957(1)~~ & ~~0.12899(90)~~ & ~~1982~~ & ~~\cite{wood:1982wood}~~ & ~~$\num{8.88e-7}\pm \num{1.4e-8}$~~ \\ 
 \hline
~~18~~ & ~~$^{40}$Ar~~ & ~~0.158051(5)~~ & ~~0.1567(16)~~ & ~~1982~~ & ~~\cite{gould:2001gould}~~ & ~~$\num{1.84e-6}\pm \num{2.2e-8}  $~~ \\ 
 \hline
 \hline
\end{tabular}
 \caption{In the last column, the values for the fractional extra dimension are shown, in agreement with the theoretical and experimental results for hydrogen \cite{Sapirstein:1981ph, Lundeen:1981zz} and hydrogen-like ions with high $Z$ compiled in \cite{Yerokhin} for the $2p_{1/2}-2s_{1/2}$ transition.}
\label{tabla1}
\end{center}
\end{table*}

\section{Conclusions}

Attempts have been made by several authors to solve the proton radius puzzle using the physics of the SM as well as alternatives beyond it. Among the authors who have explored the latter option, some have studied models that consider extra dimensions based on brane worlds where deformations of the gravitational potential would give rise to a contribution capable of explaining the discrepancy in the Lamb shift between the values obtained using theoretical expressions and the measurements obtained by experiments with muonic hydrogen.
However, the brane world approximation is not the only way to include corrections into the atomic spectra by considering additional dimensions. In this work we have studied an approximation based on the generalization of the Laplacian to $d+\varepsilon$ dimensions where $\varepsilon$ is very small. This formulation produces a contribution to the Lamb shift which is linear in the extra fractional dimension $\varepsilon$. 
We have assumed that the total discrepancy between the experimental and theoretical values in the Lamb shift is exclusively due to the effect of the extra fractional dimension, which allows us to estimate a maximum bound for the fractional extra dimension in experiments with hydrogen-like ions (electronic or muonic).


Table I compiles the results for atoms with one electron. The calculated bounds show a clear tendency to increase as Z increases. For example, we obtained $\varepsilon\sim 10^{-11}$ for hydrogen, $\varepsilon\sim 10^{-9}$ for lithium and $\varepsilon\sim 10^{-7}$ for chlorine. Furthermore, working with recent measurements in muonic hydrogen we obtained a bound of the same order of magnitude as the result with a hydrogen-like ion $Z = 16$ with an electron. The bound increases when muonic deuterium is included, which makes us think that the behavior of the fractional extra dimension as a function of A in muonic atoms is of the same type as for atoms (with electrons). This result means that future measurements of the Lamb shift in muonic atoms with different and higher Z values will be useful for detecting extra fractional dimensions and, if possible to do, experiments with tauonic atoms would be very interesting for studying new physics.

\vspace{0.2cm}
\noindent
\textbf{Acknowledgments: } We want to thank to B. Schafer and A. Muller for useful clarifications about their works, particularly with the equations they obtained which we have used in this work. Work supported by FONDECYT (Chile) under Grant No. 1180753.

\end{document}